\documentclass[amsmath, amsfonts, showpacs, preprint, prb]{revtex4}
\usepackage{graphicx}
\usepackage{epsfig}
\usepackage{bm}
\usepackage{euscript}
\usepackage{dcolumn}
\usepackage{color}
\usepackage{amsmath}
\usepackage{amssymb}

\def\Tr{\mathrm{Tr}}
\def\d{\mathrm{d}}

\begin{document}

\title{Voltage-driven quantum oscillations of conductance in graphene}

\author{V.~A.~Yampol'skii$^{1,2}$, S.~S.~Apostolov$^{1,2}$, Z.~A.~Maizelis$^{1,2}$, Alex Levchenko$^{1,3}$, and Franco Nori$^{1,4}$}

\affiliation{$^\textit{1}$ Advanced Science Institute, The
Institute of Physical and Chemical Research (RIKEN), Wako-shi,
Saitama, 351-0198, Japan\\
$^\textit{2}$ A.~Ya.~Usikov Institute for Radiophysics and
Electronics Ukrainian Academy of Sciences, 61085 Kharkov,
Ukraine\\
$^\textit{3}$ Institute for Theoretical Physics, University of
California, Santa Barbara, California, 93106, USA\\
$^\textit{4}$ Department of Physics, Center for Theoretical
Physics, Applied Physics Program, Center for the Study of Complex
Systems, University of Michigan, Ann Arbor, MI 48109-1040, USA}

\begin{abstract}
Locally-gated single-layer graphene sheets have unusual discrete
energy states inside the potential barrier induced by a
finite-width gate. These states are localized outside the Dirac
cone of continuum states and are responsible for novel quantum
transport phenomena. Specifically, the longitudinal (along the
barrier) conductance exhibits oscillations as a function of
barrier height and/or width, which are both controlled by a nearby
gate. The origin of these oscillations can be traced back to
singularities in the density of localized states. These graphene
conductance-oscillations resemble the Shubnikov-de-Haas (SdH)
magneto-oscillations; however, here these are driven by an
electric field instead of a magnetic field.
\end{abstract}

\date{\today}


\maketitle

\section{Introduction}

The unusual and rather remarkable transport properties of graphene
continue to attract considerable attention. Soon after its
experimental discovery~\cite{Novoselov-graphene}, studies found:
unconventional quantum Hall effect~\cite{IQHE}; the possibility of
testing the Klein paradox~\cite{Klein-paradox}; specular Andreev
reflection and Josephson effect~\cite{Andreev-Josephson}; new
electric field effects~\cite{Electric-field,Voltage-driven-Dos};
intriguing electron lensing~\cite{Lensing}; and other fascinating
phenomena (see, e.g., recent
papers~\cite{Review-1,Review-2,Review-3,Gusynin,williams,dicarlo,fogler,bliokh,rozhkov}
and references therein). Studies of graphene are also inspired by
their potential application in nano-electronic devices, since an
applied electric field can vary considerably the electron
concentration and have both electrons and holes as charge carriers
with high mobility.

The subject of the present study, which is a logical continuation
of recent work~\cite{Voltage-driven-Dos}, is an unusual novel
transport effect, namely, \textit{voltage-driven} quantum
oscillations in the conductance of a single-layer gated graphene.
These oscillations originate from a new type of electron states in
graphene. When a graphene sheet is subject to nearby gates, these
create an energy barrier for propagating electrons. Here we
explicitly demonstrate that, in contrast to non-relativistic
quantum mechanics, where localized states can exist only inside
quantum wells, Dirac-like \textit{relativistic} electrons in
graphene allow energy states localized \textit{within the
barrier}. We show that the energy $\varepsilon(q_{y})$ of the
localized states (versus the wave vector component $q_y$ along the
barrier) becomes \textit{non-monotonic} if
\[
V_{0}D>\pi\hbar v_{F},
\]
where $V_{0}$ and $D$ are the barrier height and width
correspondingly, and $v_{F}$ is the Fermi velocity. This produces
singularities in the density of localized states for energies
where
\[
\d\varepsilon/\d q_{y}=0.
\]
When the magnitude and/or width of the barrier changes, the
locations of the singularities move and periodically cross the
Fermi level, generating quantum oscillations in the
\textit{longitudinal} (along the barrier) conductance as well as
in the thermodynamic properties of graphene. This situation
resembles the well known physical mechanism for Shubnikov-de-Haas
(SdH) magneto-oscillations (see, e.g., Ref.~\cite{Shoenberg,gus}).
Indeed, electrons in the conduction band of a 3D metal subject to
a strong magnetic field have an equidistant discrete energy levels
(Landau levels) separated by the cyclotron energy. The
corresponding density of states has singularities at the Landau
levels. When the magnetic field is changed, the positions of the
Landau levels move and pass periodically through the Fermi energy.
As a result of this, the population of electrons at the Fermi
level also changes periodically, giving rise to the quantum
oscillations of both the transport and thermodynamic properties of
a metal. One should notice, however, a few important differences.
First, in the context of gated graphene, oscillations are induced
by the \textit{electric} field, while the corresponding SdH
oscillations are driven by a \textit{magnetic} field. Second,
localized energy states in graphene are non-equidistant and the
resulting density of states has a rather complicated energy
dependence. Thus, the corresponding oscillations in the
conductance inherit all these unusual peculiarities.

\section{Localized energy states in a barrier}

\begin{figure}
\includegraphics[width=8cm]{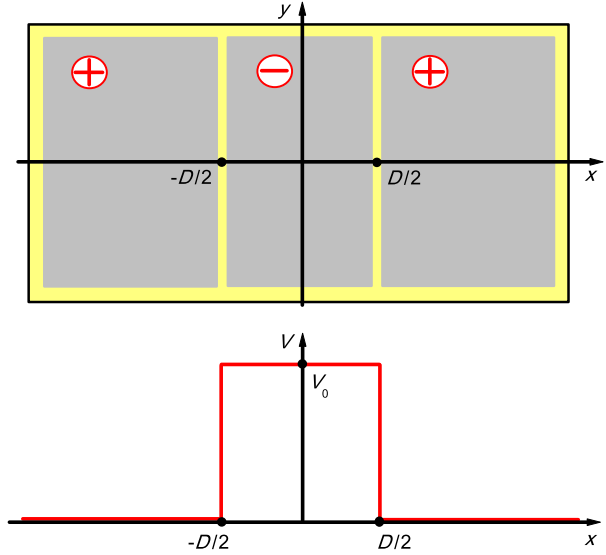} \caption{(Color online)
Schematic top view of a graphene sheet (yellow rectangle) placed
under voltage gates indicated by the grey block rectangles.
Bottom: gate-induced potential energy barrier $V(x)$ in graphene.
\label{Fig1}}
\end{figure}

The tunneling of relativistic particles in graphene
\textit{across} a finite-width potential barrier, and its
corresponding conductance, has been recently studied (see, e.g.,
Refs.~\cite{Klein-paradox,Tunnel-1,Tunnel-2,Tunnel-3,Lensing}).
Here we consider another conduction problem, namely, electron
waves that propagate \textit{strictly along} the barrier and
\textit{damp away from it}. More specifically, we consider
electron states in graphene with a potential barrier located in a
single-layer graphene occupying the $xy$-plane (see
Fig.~\ref{Fig1}). For simplicity, we assume that the barrier
$V(x)$ has sharp edges,
\begin{equation}\label{V-potential}
V(x)=\left\{\begin{array}{cr} 0, &\quad |x|>D/2\,, \\
V_{0}, &\quad |x|<D/2\,.
\end{array}\right.
\end{equation}
Electrons in monolayer graphene obey the Dirac-like equation
(hereafter $\hbar=1$),
\begin{equation}\label{Dirac-Eq}
i\frac{\partial\psi}{\partial t}=\hat{H}\psi, \quad
\hat{H}=-iv_{F}\hat{\bm{\sigma}}\cdot\bm{\nabla}+V(x)\,,
\end{equation}
where $\hat{\bm{\sigma}}=(\hat{\sigma}_{x},\hat{\sigma}_{y})$ are
the Pauli matrices. We then seek stationary spinor solutions of
the form,
\begin{equation}\label{Trial-wave}
\psi(x,y)=\psi(x)\exp(-i\varepsilon t+iq_{y}y)\,,
\end{equation}
with energy $\varepsilon$ and momentum $q_{y}$ along the barrier.
We focus on states with $|q_y|>|\kappa|\equiv
|\varepsilon|/v_{F}$. In this case, the electron waves satisfying
Eq.~\eqref{Dirac-Eq} damp away from the barrier, and the
components $\psi_1$ and $\psi_2$ of the Dirac spinor can be
written in the from
\begin{equation}\label{Psi-1}
\hskip-1.2cm \psi_{1}(x)=\left\{
\begin{array}{ll}
a\,\exp\!{\big[k_{x}(x+D/2)\big]}, &\, x<-D/2\,, \\
b\,\exp{(iq_{x}x)}\\\quad+\;c\,\exp{(-iq_{x}x)}\,, &  |x|\leq D/2\,,\\
d\, \exp\!{\big[-k_{x}(x-D/2)\big]}\,, & \,\, x>D/2\,,
\end{array}\right.
\end{equation}
\begin{equation}\label{Psi-2}
\psi_{2}(x)=\left\{
\begin{array}{ll}
a\,\frac{i\kappa}{k_{x}+q_{y}}\,\exp\!{\big[k_{x}(x+D/2)\big]}\,, &\, x<-D/2\,, \\
-b\,\exp\!{(iq_{x}x+i\theta)}\\\quad+\;c\,\exp\!{(-iq_{x}x-i\theta)}\,, & |x|\leq D/2\,,\\
-d\,\frac{i\kappa}{k_{x}-q_{y}}\,
\exp\!{\big[-k_{x}(x-D/2)\big]}\,, & \,\, x>D/2\,,
\end{array}
\right.
\end{equation}
with real $k_{x}=\big({q^{2}_{y}-\kappa^2}\big)^{1/2}$ and
$q_{x}=\big((\kappa-\mathcal{V}/D)^2-q^{2}_{y}\big)^{1/2}$. Here
$\mathcal{V}=V_{0}D/v_{F}$ is the effective barrier strength and
$\tan\theta=q_{y}/q_{x}$.

Matching the functions $\psi_{1}(x)$ and $\psi_{2}(x)$ at the points
$x=\pm D/2$, we obtain a set of four linear homogeneous algebraic
equations for the constants $a$, $b$, $c$, and $d$. Equating the
determinant of this set to zero, we obtain a dispersion relation for
the localized electron energy states,
\begin{equation}\label{Spectrum}
F(\varepsilon,q_y)\equiv\tan(q_{x}D)+\frac{k_{x}q_{x}}
{\kappa(\mathcal{V}/D-\kappa)+q^{2}_{y}}=0\,.
\end{equation}

\begin{figure}[b!]
\includegraphics[width=8cm]{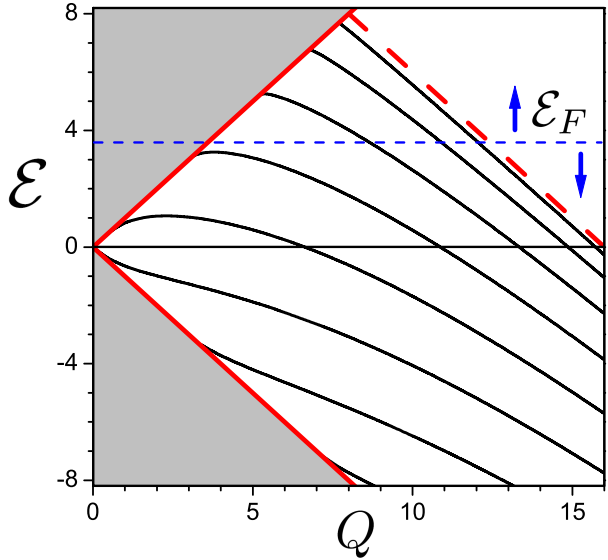} \caption{(Color online)
Electron energy spectrum in graphene obtained for positive $Q$ and
$\mathcal{V}=16$. The Dirac sea of delocalized states (continuum
spectrum) is marked by the grey regions. The branches of the
spectrum for \textit{localized} states are shown by solid black
curves between the straight solid and dashed red lines. There are
no states in the forbidden (white) regions. The increase or
decrease (schematically indicated by the upward and downward
vertical blue arrows) of the Fermi level $\mathcal{E}_F$ (marked
by the horizontal dashed blue line) results in a periodic change
in both the density of states at $\mathcal{E}=\mathcal{E}_F$, and
also in the conductance. \label{Fig2}}
\end{figure}

The spectrum of localized states in graphene
[Eq.~\eqref{Spectrum}] is shown by the solid black curves in
Fig.~\ref{Fig2}, for dimensionless variables
\begin{equation}
Q=q_{y}D,\qquad \mathcal{E}=\varepsilon D/v_F.
\end{equation}
This spectrum consists of an infinite number of branches
$\mathcal{E}_{n}(Q)$. Each of these branches starts from the lines
$\mathcal{E}=\pm|Q|$ (red solid straight lines in Fig.~\ref{Fig2})
at
\begin{equation}\label{en}
\mathcal{E}=\mathcal{V}/2-\pi^2n^2/2\mathcal{V}
\end{equation}
and tends asymptotically to the line
\[
\mathcal{E}=\mathcal{V}- Q
\]
with increasing $Q$ (dashed red line in Fig.~\ref{Fig2}).
Furthermore, a particular branch of the spectrum starts at the
point ($Q=0,\, \mathcal{E}=0$) and also tends to the line
$\mathcal{E}=\mathcal{V}-Q$.

The behavior of different branches of the spectrum depends on the
barrier strength $\mathcal{V}$. If $\mathcal{V}<\pi/2$, then all
branches satisfy $\mathcal{E}<0$. Localized states with positive
energies appear only for $\mathcal{V}>\pi/2$. When $\mathcal{V}$
increases, new branches in the spectrum with positive energies
appear. When $\mathcal{V}$ is within the interval
\[
(n+1/2)\pi<\mathcal{V}<(n+3/2)\pi,
\]
the number of branches with $\mathcal{E}>0$ is $n+1$, for
$n=1,2,3,\ldots$. It is worth emphasizing that each of the
branches with positive energy has a maximum
$\mathcal{E}^{\mathrm{max}}_{n}$ at a certain wave number
$Q=Q^{\mathrm{max}}_{n}$. Near these points the group velocity of
localized electron waves tends to zero, which resembles the
stop-light phenomena found in various media~\cite{Stop-light}. The
localized states can also be observed in graphene when a voltage
is applied to produce a potential well~\cite{gr}.

Note that defect-induced localized electron states in graphene and
the enhancement of conductivity due to an increase of the electron
density of states localized near the graphene edges were recently
reported~\cite{LDOS}. In contrast to these examples, the electron
states studied here are localized within the barrier and also these
\textit{are tunable}, i.e., the energy levels can be shifted by
charging  the barrier strength (e.g., via tuning a gate voltage).

\section{Density of localized states}

To calculate the density of electron states $\rho(\varepsilon)$,
we use the general formula
$\rho(\varepsilon)=\sum_{\alpha}\delta(\varepsilon-\varepsilon_{\alpha})$,
where the index $\alpha$ labels the quantum state and $\delta(x)$
is Dirac's delta-function. Using
\begin{equation}
\sum_{\alpha}(\ldots)=4L_{x}L_{y}(2\pi)^{-2}\int
dk_{x}dk_{y}(\ldots)
\end{equation}
for a continuum spectrum one finds the already familiar expression
\begin{equation}\label{Dos-cont}
\rho_{{_\mathrm{cont}}}(\mathcal{E})=\rho_{_0}|\mathcal{E}|, \quad
\rho_{_0}=\frac{2L_{x}L_{y}}{\pi v_{F}D}\,,
\end{equation}
where $L_x$ and $L_y$ are the lengths of the graphene sheet in the
$x$ and $y$ directions, respectively. For localized energy states,
we obtain:
\begin{equation}\label{Dos-loc}
\rho_{{_\mathrm{loc}}}(\mathcal{E})=2\rho_{_0}\frac{D}{L_x}\sum_{n}
\left|\frac{d\mathcal{E}_{n}(Q)}{dQ}\right|^{-1}_{\mathcal{E}_{n}(Q)=\mathcal{E}}\,,
\end{equation}
where $n$ runs over the positive roots of the equation
$\mathcal{E}(Q)=\mathcal{E}$. The function
$\rho_{{_\mathrm{loc}}}(\mathcal{E})$ exhibits two types of
peculiarities. First, increasing $\mathcal{E}$, the jumps or steps
(each of magnitude $2D/L_x$) in
$\rho_{{_\mathrm{loc}}}(\mathcal{E})/\rho_{_0}$ occur at the
points, given by Eq.~(\ref{en}),
where new branches of the spectrum arise or disappear. More
importantly, \textit{singularities} are observed when
$\mathcal{E}=\mathcal{E}^{\mathrm{max}}_{n}$, where
$|\d\mathcal{E}/\d Q|^{-1}$ in Eq.~\eqref{Dos-loc} diverges.

\begin{figure}
\includegraphics[width=8cm]{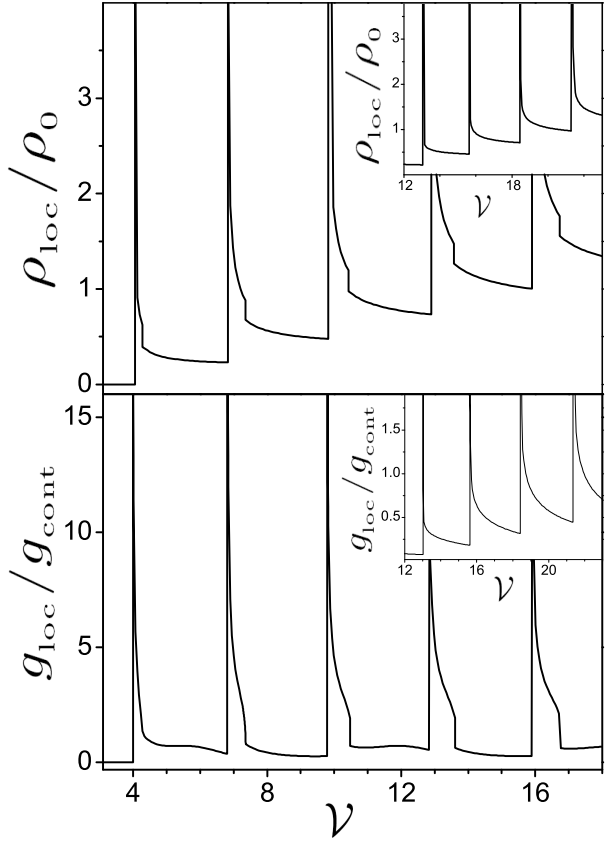} \caption{Dimensionless
oscillating parts of the density of states
$\rho_{{_\mathrm{loc}}}/\rho_{_0}$ at the Fermi level (upper
panel) and conductance $ g_{{_\mathrm{loc}}}/g_{{_\mathrm{cont}}}$
(lower panel) versus the strength $\mathcal{V}$ of the potential
barrier for $D/L_x=0.1$, $\mathcal{E}_{F}=1$ (main panels) and
$\mathcal{E}_{F}=5$ (insets). The total conductance is
$g=g_{{_\mathrm{cont}}}+ g_{{_\mathrm{loc}}}$. \label{Fig3}}
\end{figure}

The locations of the singularities \textit{shift} when changing
the barrier strength $\mathcal{V}$. Therefore, they periodically
cross the Fermi level $\mathcal{E}_{F}$. This produces quantum
oscillations in the density of states at the Fermi energy, which
are seen in the upper panel of Fig.~\ref{Fig3}, showing
$\rho_{{_\mathrm{loc}}}(\mathcal{E})/\rho_{_0}$ versus the
effective barrier strength $\mathcal{V}$.

\section{Kubo formula and conductance}

\subsection{Kubo expression for the conductance in graphene}

When studying transport, within linear response theory, one
usually starts from the current-response function,
\begin{equation}\label{K}
K_{\mu\nu}(\mathbf{x},\mathbf{x}')=-i\vartheta(t-t')\Tr\left\{\hat{\varrho}
\big[\hat{\mathbf{j}}^{H}_{\mu}(\mathbf{x}),
\hat{\mathbf{j}}^{H}_{\nu}(\mathbf{x}')\big]\right\}\,,
\end{equation}
where $\mathbf{x}=(\mathbf{r},t)$, $\vartheta(t)$ is the Heaviside
step-function, $\hat{\varrho}$ is the equilibrium density matrix,
and
\[
\hat{\mathbf{j}}^{H}_{\mu}(\mathbf{r},t)=
\exp{(i\hat{H}t)}\hat{\mathbf{j}}_{\mu}(\mathbf{r})\exp{(-i\hat{H}t)}
\]
is the current operator in the Heisenberg representation with the
Hamiltonian taken from Eq.~\eqref{Dirac-Eq}, and where $[\ldots,
\ldots]$ stands for the commutator. For electrons with a linear
Dirac spectrum, one finds
\begin{equation}
\hat{\mathbf{j}}_{\mu}(\mathbf{r})=ev_{F}\hat{\psi}^{\dag}(\mathbf{r})
\hat{\sigma}_{\mu}\hat{\psi}(\mathbf{r}).
\end{equation}
Equation \eqref{K} is used to define the frequency-dependent
linear conductance as
\begin{equation}\label{g-def}
g_{\mu\nu}(\omega)=\Re\frac{i}{\omega L_{\mu}L_{\nu}}\iint
\d\mathbf{r}\;\d\mathbf{r}'\,
K_{\mu\nu}(\mathbf{r},\mathbf{r}';\omega)\,.
\end{equation}
Here $\Re$ stands for the real part of a complex number.

We expand the fermionic field operator $\hat{\psi}(\mathbf{r},t)$
in terms of exact eigenfunctions [Eq.~\eqref{Trial-wave}], namely,
\begin{equation}
\hat{\psi}(\mathbf{r},t)=\sum_{\alpha}\psi_{\alpha}(\mathbf{r})
\exp{(-i\epsilon_{\alpha}t\hat{a}_{\alpha})},
\end{equation}
and then perform quantum averaging in Eq.~\eqref{K} with the help
of Wick's theorem and the relation
$\Tr\big\{\hat{\varrho}\,\hat{a}^{\dag}_{\alpha}\hat{a}_{\beta}\big\}=
\delta_{\alpha\beta}f(\epsilon_{\alpha})$, where
\[
f(\varepsilon)=1/[\exp[(\varepsilon-\varepsilon_{F})/T]+1]
\]
is the Fermi occupation function. Performing a Fourier transform
and using
\begin{equation}
\Re[i/(\varepsilon-\varepsilon'+\omega+i0)]=\pi\delta(\varepsilon-\varepsilon'+\omega)
\end{equation}
Eq.~\eqref{g-def}, reduces to
\begin{eqnarray}
g_{\mu\nu}(\omega)=\frac{\pi(ev_{F})^2}{L_{\mu}L_{\nu}}\int^{+\infty}_{-\infty}
\d\varepsilon\,
\frac{f(\varepsilon_{+})-f(\varepsilon_{-})}{\omega}\nonumber\\\times
\Tr\left\{\hat{\sigma}_{\mu}
\delta(\epsilon_{+}-\hat{H})_{\mathbf{rr}'}\hat{\sigma}_{\nu}
\delta(\varepsilon_{-}-\hat{H})_{\mathbf{r}'\mathbf{r}}\right\}\,,
\end{eqnarray}
where $\varepsilon_{\pm}=\varepsilon\pm\omega/2$ and the trace
incorporates spatial integrations. The operator delta-functions
can be directly related to the single-particle Green's functions
\[
\hat{G}_{\varepsilon}(\mathbf{r},\mathbf{r}')=
\langle\mathbf{r}|(\varepsilon-\hat{H})^{-1}|\mathbf{r}'\rangle
\]
according to
\begin{equation}
\delta(\varepsilon-\hat{H})_{\mathbf{rr}'}=\frac{1}{2\pi
i}\big[\hat{G}^{a}_{\varepsilon}(\mathbf{r},\mathbf{r}')-
\hat{G}^{r}_{\varepsilon}(\mathbf{r},\mathbf{r}')\big]\,,
\end{equation}
where the superscript $a/r$ stands for the advanced/retarded
component, respectively. As a result, one finds for the
conductance
\begin{eqnarray}
&& \hskip-0.5cm g_{\mu\nu}(\omega)=\frac{(ev_{F})^2}{4\pi
L_{\mu}L_{\nu}}\int^{+\infty}_{-\infty} \d\varepsilon\,
\frac{f(\varepsilon_{+})-f(\varepsilon_{-})}{\omega} \\\times\,
&&\hskip-0.5cm \Tr\!\left\{\!\hat{\sigma}_{\mu}
\big[\hat{G}^{a}_{\varepsilon_{+}}\!(\mathbf{r},\mathbf{r}')\!-\!
\hat{G}^{r}_{\varepsilon_{+}}\!(\mathbf{r},\mathbf{r}')\big]
\hat\sigma_{\nu}
\big[\hat{G}^{r}_{\varepsilon_{\!-}}\!(\mathbf{r}',\mathbf{r})\!-\!
\hat{G}^{a}_{\varepsilon_{\!-}}\!(\mathbf{r}',\mathbf{r})\big]
\!\right\}\nonumber.
\end{eqnarray}
Next we incorporate disorder by introducing the one-particle
scattering time $\tau$, for Dirac fermions, into the Green's
function,
\begin{equation}\label{G-average}
\langle\hat{G}^{r/a}_{\varepsilon}\rangle_{\mathrm{dis}}\approx(\varepsilon-\hat{H}\pm
i/\tau)^{-1}\,,
\end{equation}
which enters through the imaginary-part of the corresponding
self-energy. The subindex ``dis'' refers to disorder. Furthermore,
we factorize the average of the product of two Green's functions
by the product of their averages,
\begin{equation}
\langle\hat{G}^{r}_{\varepsilon_{+}}\hat{G}^{a}_{\varepsilon_{-}}\rangle_{\mathrm{dis}}\approx
\langle\hat{G}^{r}_{\varepsilon_{+}}\rangle_{\mathrm{dis}}
\langle\hat{G}^{a}_{\varepsilon_{-}}\rangle_{\mathrm{dis}}\,.
\end{equation}
This assumption should be valid for weak disorder and together
with Eq.~\eqref{G-average} is equivalent to the self-consistent
Born approximation.

\subsection{Conductance along the barrier}

We now focus on the \textit{along-the-barrier} ($\mu=\nu=y$)
conductance for the geometry shown in Fig.~\ref{Fig1}. At zero
temperature, $T\to 0$, when
$f(\varepsilon)=\vartheta(\varepsilon_{F}-\varepsilon)$ and the
$\varepsilon$ integration is bounded by the frequency $\omega$,
for the average dc-conductance $g\equiv\langle
g_{yy}(\omega\to0)\rangle_{\mathrm{dis}}$ we find (per spin and
per valley):
\begin{equation}
g=g_{{_\mathrm{cont}}}+ g_{{_\mathrm{loc}}}\,.
\end{equation}
The first contribution $g_{{_\mathrm{cont}}}$ here comes from the
extended electron energy states with corresponding density of
states taken from Eq.~\eqref{Dos-cont}, and reads explicitly (now
keeping $\hbar$) as
\begin{equation}\label{g-cont}
g_{{_\mathrm{cont}}}=\frac{\pi e^2}{16\hbar}\frac{L_x}{L_y}\left[
\varepsilon_{F}\tau+\frac{1}{\pi}
\left(1-\varepsilon_{F}\tau\arctan\frac{1}{\varepsilon_{F}\tau}\right)\right]\,.
\end{equation}
At the neutrality point, $\varepsilon_{F}=0$, from
Eq.~\eqref{g-cont} one recovers a universal (i.e., scattering time
$\tau$-independent) result
$g_{{_\mathrm{cont}}}=\sigma_{\mathrm{min}}(L_{x}/L_{y})$, where
$\sigma_{\mathrm{min}}=(\pi/8)(e^2/h)$ is the minimal
conductivity, which received considerable attention in a number of
recent studies (e.g., Refs.~\cite{Tunnel-2,Minimal}). Away from
the neutrality point, the conductance growths linearly with the
Fermi energy,
\begin{equation}
g_{{_\mathrm{cont}}}=(\pi
e^2/16\hbar)(L_x/L_y)\varepsilon_{F}\tau.
\end{equation}

The novel result of the present study is the \textit{oscillatory}
part $ g_{{_\mathrm{loc}}}$, which originates from the electron
states localized within the barrier. It can be expressed, with the
help of Eq.~\eqref{Dos-loc}, as follows:
\begin{equation}\label{g-loc}
 g_{{_\mathrm{loc}}}=\frac{2 e^2}{\hbar}\frac{D}{L_{y}}\sum_{n}\int^{\infty}_{0}\!\!
\d\mathcal{E}\, \left|\frac{\d
Q}{\d\mathcal{E}_{n}}\right|_{\mathcal{E}_{n}=\mathcal{E}}
\frac{M(\mathcal{E})\eta^2}{\big[(\mathcal{E}-\mathcal{E}_{F})^2+\eta^2\big]^2},
\end{equation}
where
\[
M(\mathcal{E})=\left|\int\frac{\d
x}{D}\psi^{*}_{\alpha}(x)\hat{\sigma}_{y}\psi_{\alpha}(x)\right|^2
\]
is the matrix element constructed from the wave-functions of
localized states, Eq.~\eqref{Psi-1}-\eqref{Psi-2}, and
$\eta=D/v_{F}\tau$. The remaining integration in Eq.~\eqref{g-loc}
is simplified realizing that everywhere away from the integrable
square-root singularities of $|\d Q/\d\mathcal{E}_{n}|$, the
$\eta$-dependent function is peaked at the Fermi energy, whereas
$M(\mathcal{E})$ is smooth. Thus, one finally finds,
\begin{equation}\label{g-over-g}
\frac{ g_{{_\mathrm{loc}}}(\mathcal{V},\mathcal{E}_{F})}
{g_{{_\mathrm{cont}}}}=\frac{16}{\mathcal{E}_{F}}\frac{D}{L_{x}}M(\mathcal{E}_{F})
\sum_{n}\left|\frac{\d Q}{\d
\mathcal{E}_{n}}\right|_{\mathcal{E}_{n}=\mathcal{E}_{F}}\,,
\end{equation}
where the conductance $ g_{{_\mathrm{loc}}}$ is normalized to its
continuous part taken away from the neutrality point, namely,
where $g_{{_\mathrm{cont}}}\propto\tau\varepsilon_{F}$. Note that
\[
\mathcal{E}_{F}=\varepsilon_{F}\frac{D}{v_F}.
\]
The derivative entering Eq.~\eqref{g-over-g} can be
calculated with the help of the dispersion equation
\eqref{Spectrum} as $(\d Q/\d\mathcal{E})=-(\d
F/\d\mathcal{E})/(\d F/\d Q)$, and reads
\begin{equation}
\frac{\d
Q}{\d\mathcal{E}}=Q\frac{\mathcal{V}-2\mathcal{E}+(\mathcal{V}-\mathcal{E})\sqrt{Q^2-\mathcal{E}^2}}
{(\mathcal{V}-\mathcal{E})\mathcal{E}-Q^2-Q^2\sqrt{Q^2-\mathcal{E}^2}}\,.
\end{equation}
The oscillatory nature of
$g_{{_\mathrm{loc}}}(\mathcal{V},\mathcal{E}_{F})$ is illustrated
in the lower panel of Fig.~\ref{Fig3}. The essential observation,
which follows from Eq.~\eqref{g-over-g}, is that the longitudinal
conductance traces the peculiarities in the density of localized
states and opens a direct way for their experimental observation.
It is also worth mentioning that close to the singularity of $\d
Q/\d \mathcal{E}$, meaning
$|\mathcal{E}_{n}-\mathcal{E}_{F}|\lesssim\eta$, the conductance
correction is regularized by the finite width of the
$\eta$-Lorenzian under the integral of Eq.~\eqref{g-loc}.

\begin{figure}
\includegraphics[width=8cm]{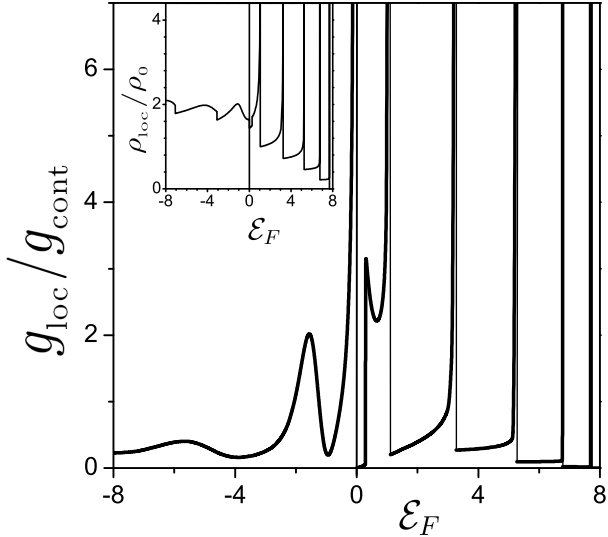} \caption{Dimensionless
oscillating parts of the density of states
$\rho_{{_\mathrm{loc}}}/\rho_{_0}$ at the Fermi level (inset) and
conductance $ g_{{_\mathrm{loc}}}/g_{{_\mathrm{cont}}}$ (main
panel) versus the Fermi energy $\mathcal{E}_{F}$, for $D/L_x=0.1$,
and $\mathcal{V}=16$. \label{Fig4}}
\end{figure}

Varying the concentration of free particles with constant barrier
strength, one can again observe oscillations in the density of
states (see the inset of Fig.~\ref{Fig4}). Thus, the part of the
conductance originated from the localized states, also oscillates
with the change of the Fermi energy (see main panel of
Fig.~\ref{Fig4}).

\section{Conclusions}

In summary, we predict a novel type of conductance oscillations in
locally-gated single-layer graphene, which are related to the
unusual electron states localized within a potential barrier. When
the barrier height and/or width is varied, localized levels
periodically cross the Fermi energy, inducing modulations in the
density of states. The latter translates into unusual quantum
oscillations of the conductance. These electric-field-driven
quantum oscillations are similar to the Shubnikov-de-Haas
oscillations which are produced in metals and semiconductors when
changing the external magnetic field.

\begin{acknowledgements}

We gratefully acknowledge partial support from the National
Security Agency (NSA), Laboratory of Physical Sciences (LPS), Army
Research Office (ARO), National Science Foundation (NSF) grant No.
EIA-0130383, JSPS-RFBR 06-02-91200, and Core-to-Core (CTC) program
supported by Japan Society for Promotion of Science (JSPS). A.L.
acknowledges partial support from the National Science Foundation
under Grant No. NSF PHY05-51164.

\end{acknowledgements}


 \end{document}